\documentclass[10pt,preprint]{aastex}

\shorttitle{Hierarchy of Timescales of QPP} \shortauthors{Tan et
al.}

\begin{document}


\title{Broadband Radio Spectral Observations of Solar Eclipse on 2008-08-01 and Implications on the Quiet Sun Atmospheric Model}
\author{Baolin Tan\altaffilmark{1}, Yihua Yan, Yin Zhang,
Chengmin Tan, Jing Huang, Yuying Liu, Qijun Fu, ZhiJun Chen,Fei
Liu, Linjie Chen and Guoshu Ji} \affil{Key Laboratory of Solar
Activity, National Astronomical Observatories
\\ Chinese Academy of Sciences, China} \email{bltan@nao.cas.cn} Received July 15, 2009; accepted July 18, 2009; published online September 18, 2009 \\
$^\dag$Corresponding author (email: bltan@nao.cas.cn)\\
\\doi:10.1007/s11433-009-0230-y, published on Sci. China Ser. G,
2009, Vol.52, page 1765-1772

\begin{abstract}

Based on the joint-observations of the radio broadband spectral
emissions of solar eclipse on August 1, 2008 at Jiuquan (total
eclipse) and Huairou (partial eclipse) at the frequencies of 2.00
-- 5.60 GHz (Jiuquan), 2.60 -- 3.80 GHZ (Chinese solar broadband
radiospectrometer, SBRS/Huairou), and 5.20 -- 7.60 GHz
(SBRS/Huairou), the authors assemble a successive series of
broadband spectrum with a frequency of 2.60 -- 7.60 GHz to observe
the solar eclipse synchronously. This is the first attempt to
analyze the solar eclipse radio emission under the two telescopes
located at different places with broadband frequencies in the
periods of total and partial eclipse. With these analyses, the
authors made a semiempirical model of the coronal plasma density
of the quiet Sun, which can be expressed as $n_{e}\simeq
1.42\times10^{9}(r^{-2}+1.93r^{-5}),(cm^{-3})$ in the space range
of $r=1.039 - 1.212 R_{\odot}$, and made a comparison with the
classic model.

\end{abstract}


\section{Introduction}

Without the high-resolution solar radioheliograph, the solar radio
eclipse observations may be the best way to investigate the
spatial structure of the solar radio emission. During the minimum
phase of solar cycle, the influence of the active regions is very
weak, we may study the quiet solar coronal atmospheric model by
means of the radio eclipse observations$^{[1]}$.

On August 1, 2008, a grand solar total eclipse could occurred on
the regions of northern Europe and northwestern China. In Jiuquan
of Gansu province, people could observe the whole course of
eclipse happening. In Beijing, people could also watch the partial
solar eclipse clearly. Around this event, the Sun was in a quiet
phase between solar cycle 23 and 24. There was no solar flare, no
corona mass ejection (CME), no sunspot, and no any active regions.
The solar disk looked like a fairly clean plate. Such total
eclipse might provide a god-given opportunity to test the quiet
Sun atmospheric models by means of radio observations. During this
solar eclipse, we made an extensive broadband spectral radio
joint-observations at Jiuquan and Beijing Huairou. Section 2
introduces the observed instruments and the related data
processing. Section 3 is a deduction of the quiet Sun atmospheric
model from the above data. And some discussions and conclusions
make up the final part of the paper.

\section{Observations and Data Processing}

The joint-observation included a solar total eclipse and partial
eclipse observations. The solar total eclipse observations were
carried out at Jiuquan with a broadband spectrometer (hereafter,
EcBS/Jiuquan) in the frequency range of 2.00 -- 5.60 GHz$^{[2]}$.
However, because of instrument problems, we only obtained a
segment of valid data in the frequency of 3.60 -- 5.60 GHz. The
telescope was a newly updated instrument manufactured just before
July of 2008 with an aperture of 2.40 m, not of very good
performance. The status of the telescope tracking the Sun was
recorded by a small CCD camera, connected to a small optical
telescope and a computer. They were in-phase with the radio
receiver by an GPS system. The axes of the CCD camera and the
radio telescope were parallel to each other. The time cadence was
about 10 pictures per second. Figure 1 presents the results of the
CCD monitoring and the occulted information at three different
times. The time cadence of the recorded radio data was 7.5 ms.
This means that, in each segment of 7.5 ms the receiver would do a
whole scan in the frequency range of 2.0 -- 5.60 GHz and recorded
them, and it makes a recorded unit. Each recorded file includes 4
MB data in about 8000 recorded units. However, as a result of the
missing and disordered data, each recorded file included actually
about 6000 -- 7000 valid recorded units. The time length in each
recorded file was about 3.0 -- 7.00 s with the averaged value
about 5.59 s. As the magnitude of the fluctuation among the
recorded units was very strong, we might integrate them into each
recorded file. As for the frequency bandwidth, we might  integrate
them into 28 MHz at each channel. The solar radio partial eclipse
observation is made by the Chinese solar broadband
radiospectrometer (SBRS/Huairou) in Beijing Huairou in the valid
frequency range of 2.60 -- 3.80 GHz and 5.20 -- 7.60
GHz$^{[3,4]}$. The time cadence of the recorded data was
integrated into 0.2 s, and the frequency bandwidth at each channel
was 10 MHZ in the range of 2.60 -- 3.80 GHz and 20 MHz in the
range of 5.20 -- 7.60 GHz. The performance of SBRS/Huairou was in
good status.

As for the solar total eclipse at Jiuquan, the first optical
contact was at 10$^{h}$16$^{m}$36.5$^{s}$ UT with the solar
photospheric radius 945.5 arcsecond and the lunar radius 978.1
arcsecond, and the totality began at 11$^{h}$13$^{m}$41.4$^{s}$ UT
and ended at 11$^{h}$15$^{m}$21.8 UT. The maximum phase was at
11$^{h}$14$^{m}$31.7$^{s}$ UT, the eclipse factor was 1.034477,
the solar photospheric radius is 945.5 arcsecond, the lunar radius
is 978.1 arcsecond, and the distance between the solar disk center
and the lunar center was 12.95 arcsecond. The last contact was at
12$^{h}$08$^{m}$24.0$^{s}$ UT with an altazimuth low to 4.4
degree. However, as the antenna could not follow the Sun's track
after 11$^{h}$45$^{m}$ UT, we could not get the radio observations
after that time. As for the partial eclipse at Huairou, Beijing,
the first optical contact was at 10$^{h}$16$^{m}$36.1$^{s}$ UT
with the solar photospheric radius 945.5 arcsecond and the lunar
radius 977.7 arcsecond. The maximum phase was at
11$^{h}$09$^{m}$23.2$^{s}$ UT with an eclipse factor of 0.910905,
and with the solar photospheric radius 945.5 arcsecond and the
lunar radius 974.7 arcsecond, and the distance between the solar
disk center and the lunar center 190.04 arcsecond. The last
contact was at 11$^{h}$59$^{m}$11.3$^{s}$ UT, much after the
sunset (The corresponding sunset time at Huairou Station was about
11$^{h}$28$^{m}$ UT). So, we could not get the observation data
about the partial eclipse after the maximum phase.

The calibration and data processing method follow the Ref.[5 and
6]. Before and after the period of eclipse, the calibrating
observations are made and and the emission measure of the sky
background along the Sun's routine corresponding to the solar
eclipse is taken. As the Sun was in its quiet phase, the
polarization of radio emission was very weak. It is unimportant to
differentiate the right circular polarized emission (RCP) and the
left circular polarized emission (LCP). This work adopted the
total emission intensity (RCP+LCP). As the instrument of
SBRS/Huairou was aging, while the telescope of EcBS/Jiuquan was a
new instrument, the signal fluctuation was fairly strong. We make
a broad averaging with the data. So our result is smoothed one (as
for the data of EcBS/Jiuquan, we make averaging in each recorded
file, so the integrated time is about 3 -- 7 s with the mean value
of 5.59 s; as for the data of SBRS/Huairou, we make averaging in
each segment of 3.2 s). At the same time, as the period of the
eclipse was very close to the sunset, it's hard to remove the
disturbances coming from the subaerial factors.

\begin{figure}
\begin{center}
  \includegraphics[width=2.2cm]{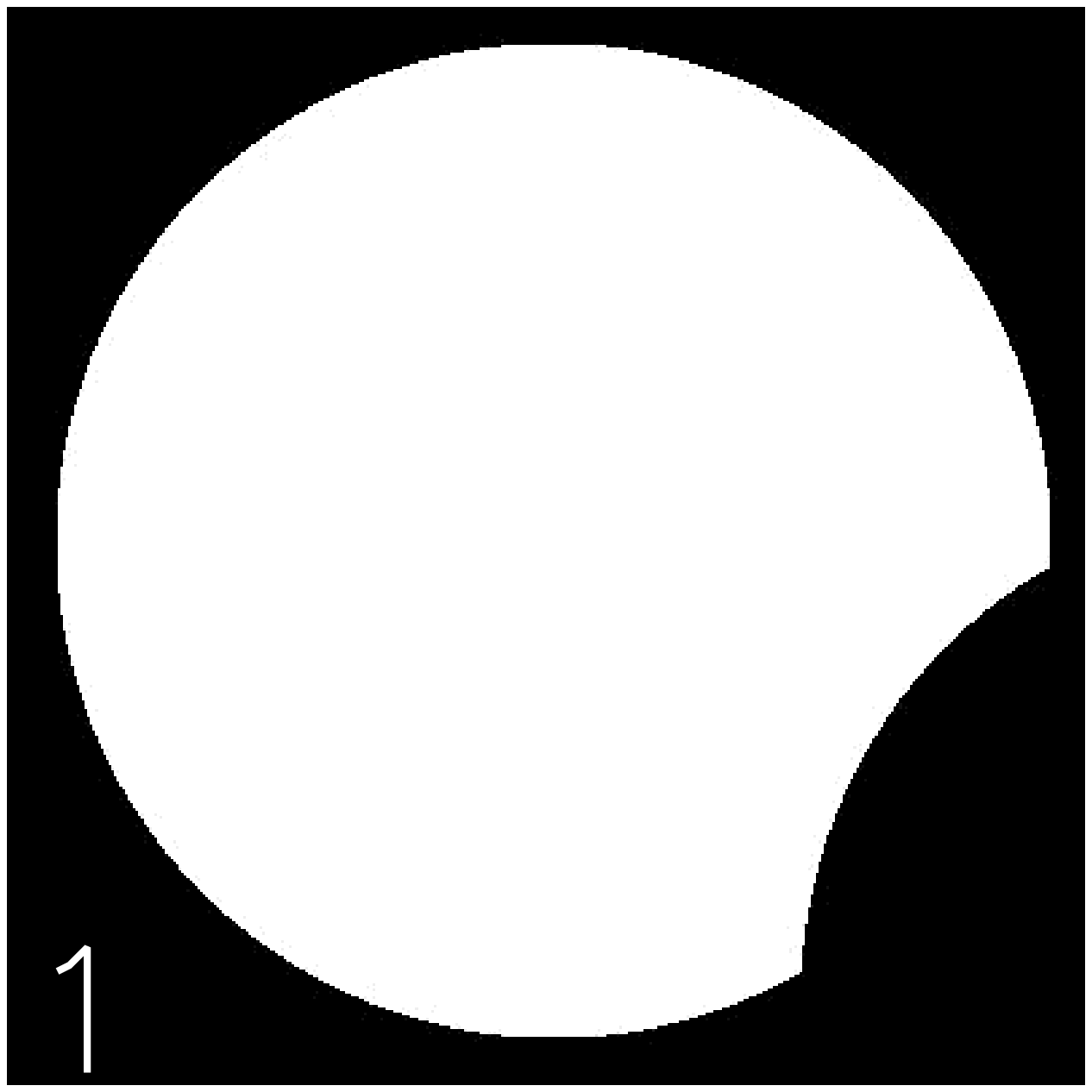}
  \includegraphics[width=2.2cm]{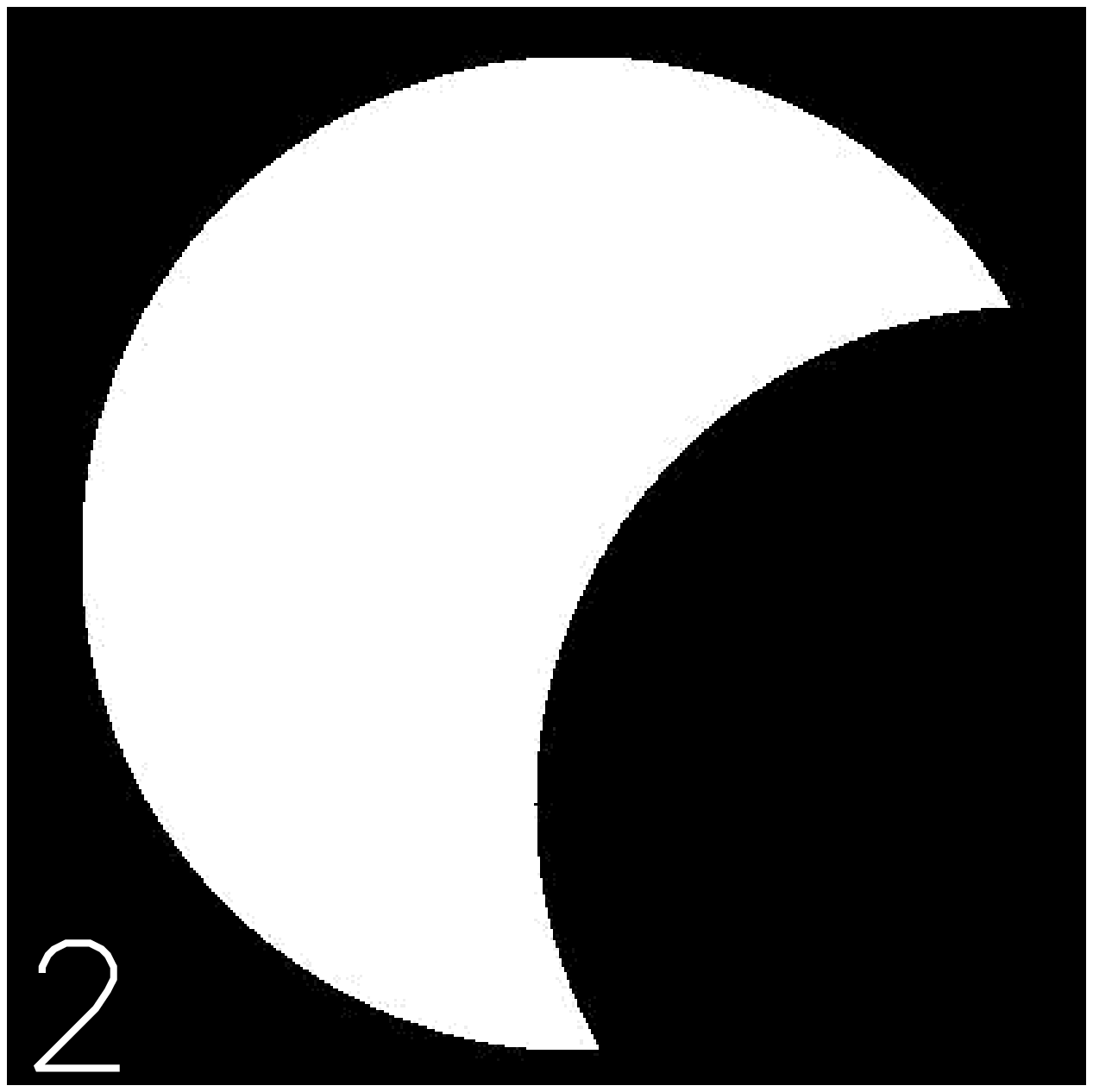}
  \includegraphics[width=2.2cm]{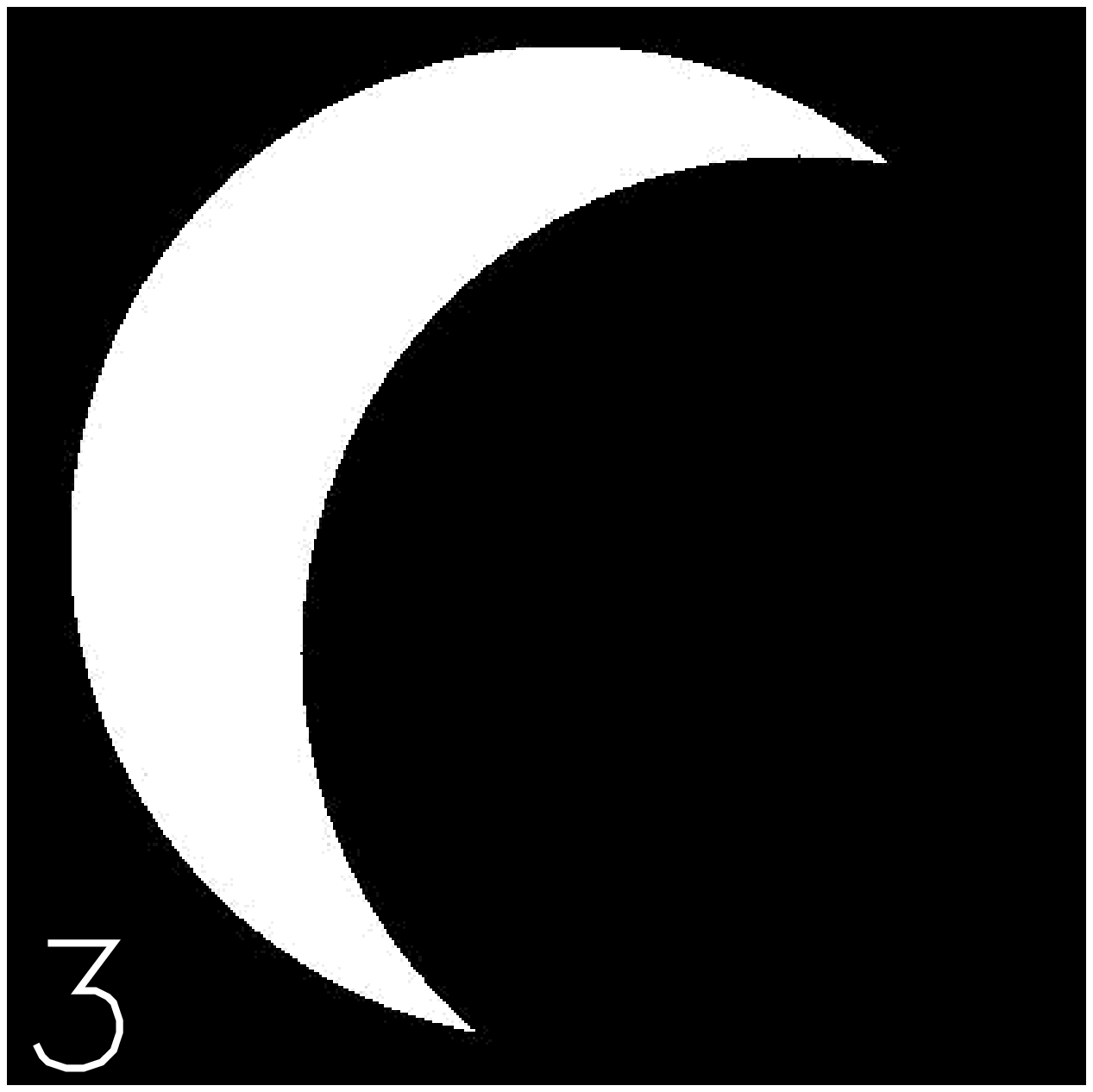}
  \\\includegraphics[width=5.4cm]{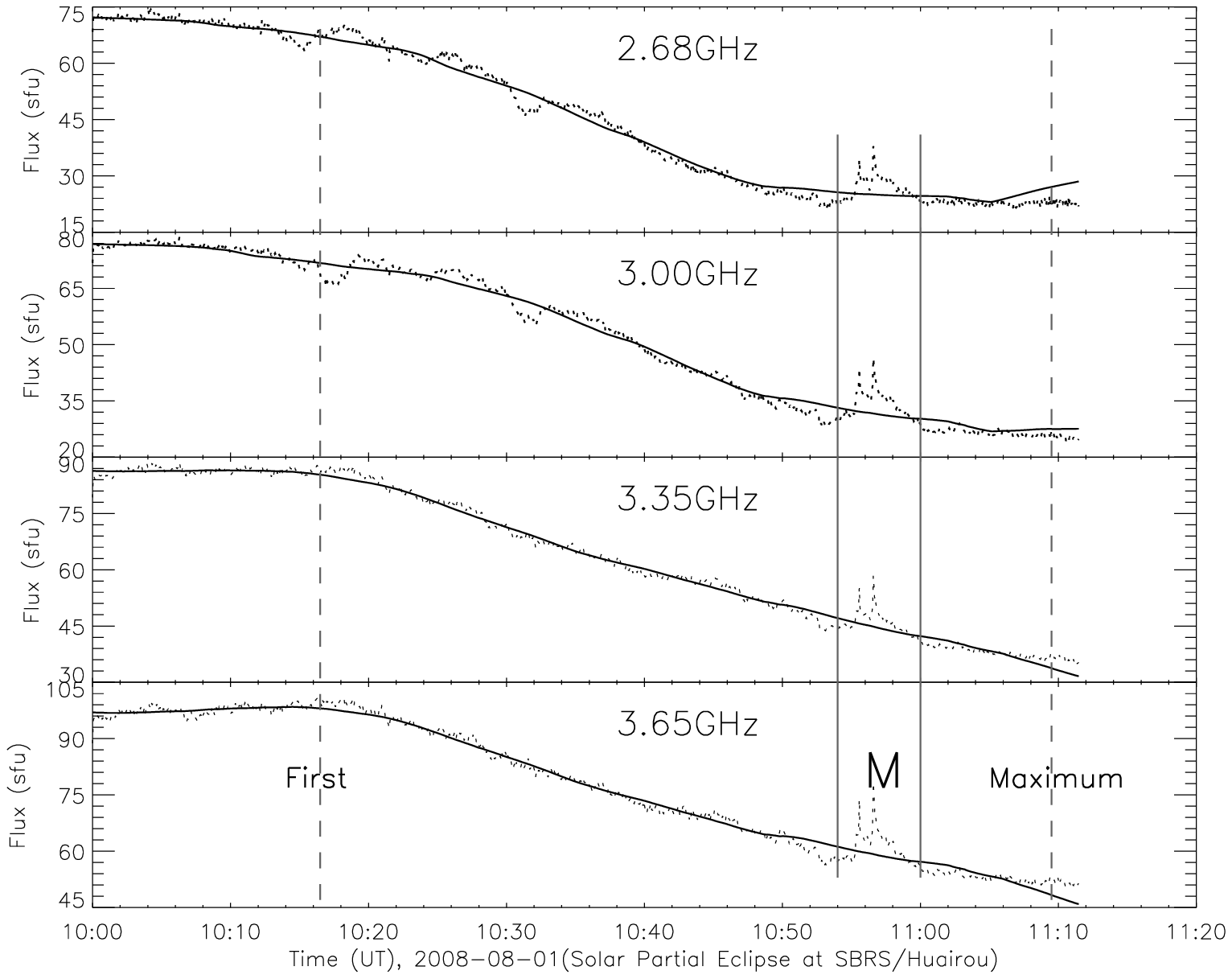}
  \includegraphics[width=5.4cm]{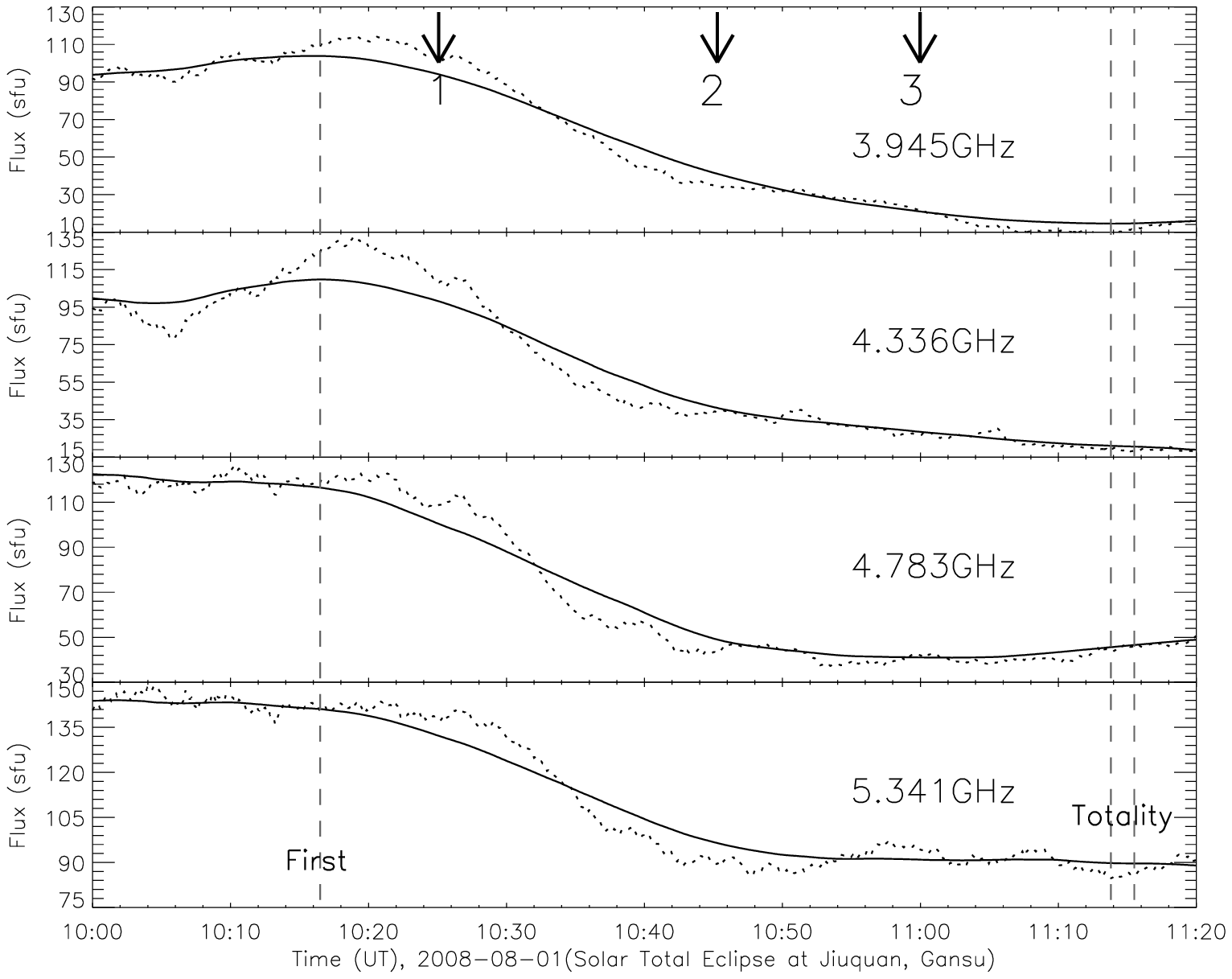}
  \includegraphics[width=5.4cm]{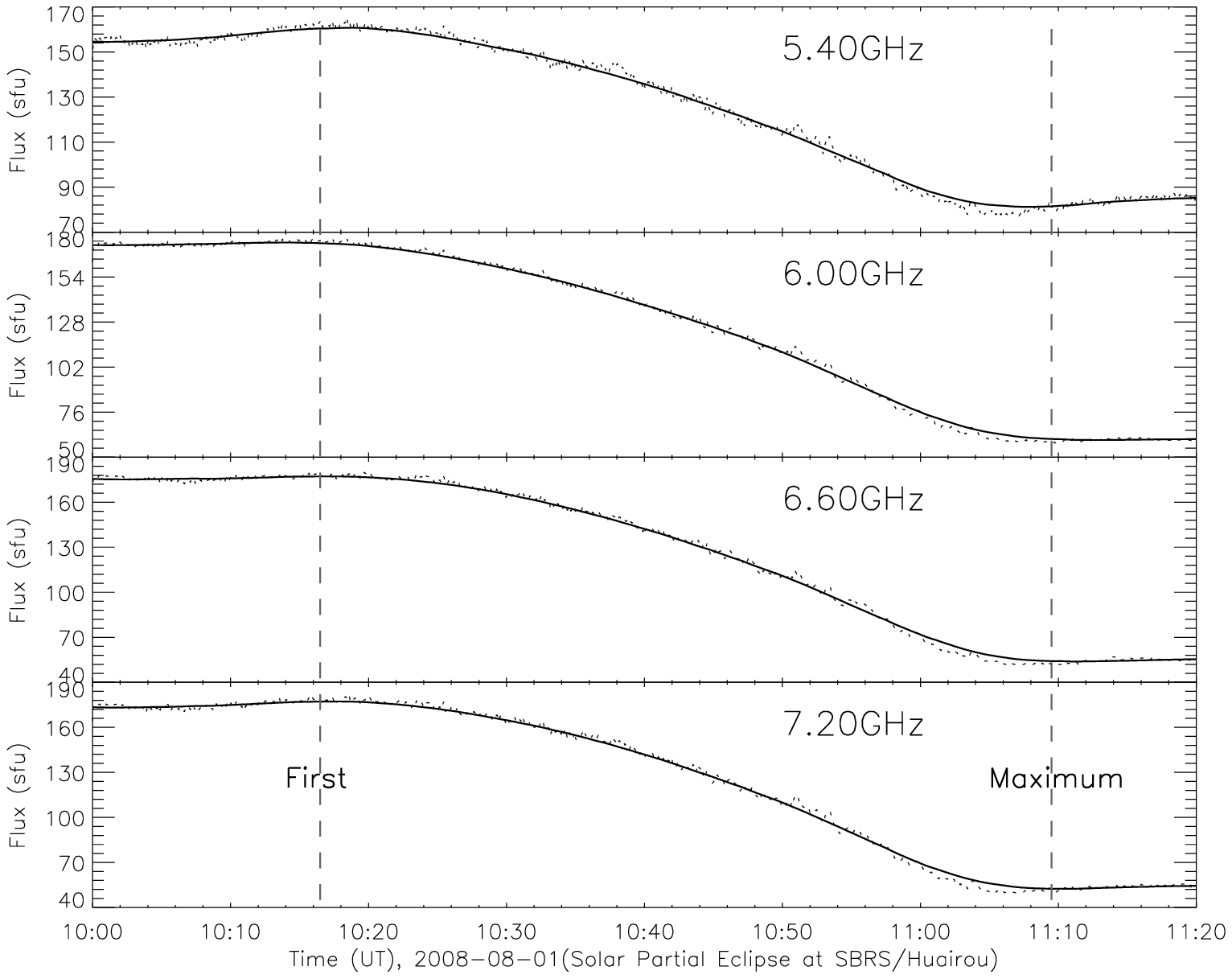}
   \caption{Solar radio eclipse curves of August 1, 2008. Left: 2.60 -- 3.80 GHz, partial eclipse, SBRS/Huairou, Beijing, the region
   marked with \textbf{M} between two vertical solid lines is most likely to be a disturbance from some unknown terrestrial factor;
   Middle: 3.80 -- 5.20 GHz, total eclipse, Jiuquan, Gansu, the upper small figures are CCD monitoring observations, from left to right
   (marked as 1, 2, 3 corresponding to the arrows with the same numbers), the time is at 10:25 UT, 10:45 UT, and 11:00 UT,
   respectively; Right: 5.20 -- 7.60 GHz, partial eclipse, SBRS/Huairou, Beijing.}
\end{center}
\end{figure}

In this work we mainly focus on the processes occurring at around
the first radio contact of the Moon's limb to the solar disk. We
select three segments of frequency to form a successive broadband
observations with frequency of 2.60 -- 7.60 GHz. The first (2.60
-- 3.80 GHz) and third (5.20 -- 7.60 GHz) segment were obtained at
SBRS/Huairou, while the second segment (3.80 -- 5.60 GHz) was
observed by EcBS/Jiuquan. Figure 1 shows the solar radio eclipse
curves which includes the Jiuquan total eclipse (left and right
panels) and the Huairou partial eclipse (middle panel). The region
marked with \textbf{M} between two vertical solid lines in the
left panel is most likely to be a disturbance from some unknown
terrestrial factor. Figure 2 presents the solar radio eclipse
slope curves one by one with Figure 1. The horizontal dashed lines
in the left panels indicate the zero level of the slopes. The
vertical solid lines in the left and right panels indicate the
time of the first optical contact and the maximum phase of the
partial eclipse, respectively. While the vertical solid lines in
the middle panels indicate the time of the first optical contact
and the totality phase of the total eclipse, respectively. All of
the solid thick curves in Figure 1 and Figure 2 are smoothed in
brand slide windows which reflect the main tendency of the solar
radio eclipse curves and the radio eclipse slope curves.

\begin{figure}
\begin{center}
  \includegraphics[width=5.4cm]{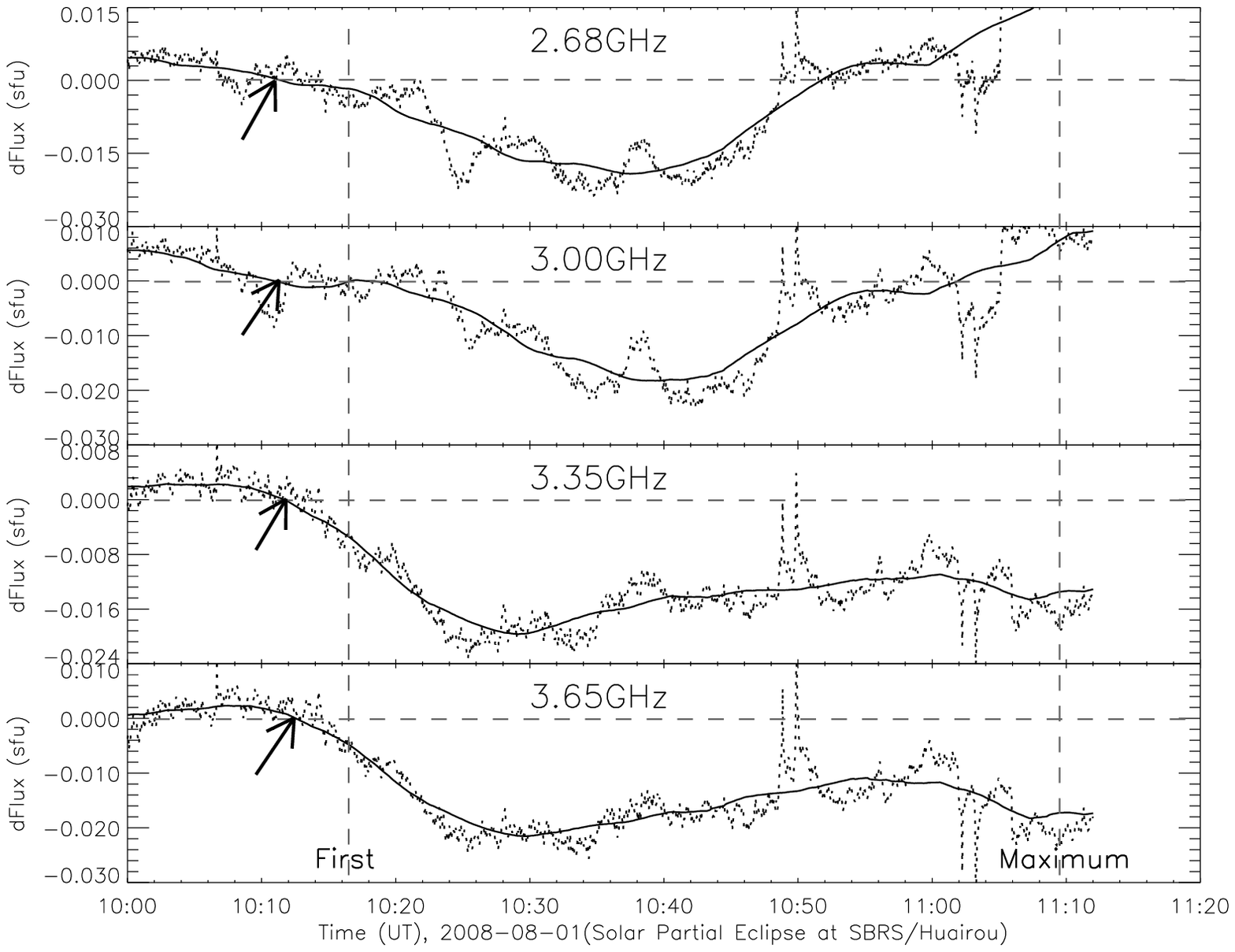}
  \includegraphics[width=5.4cm]{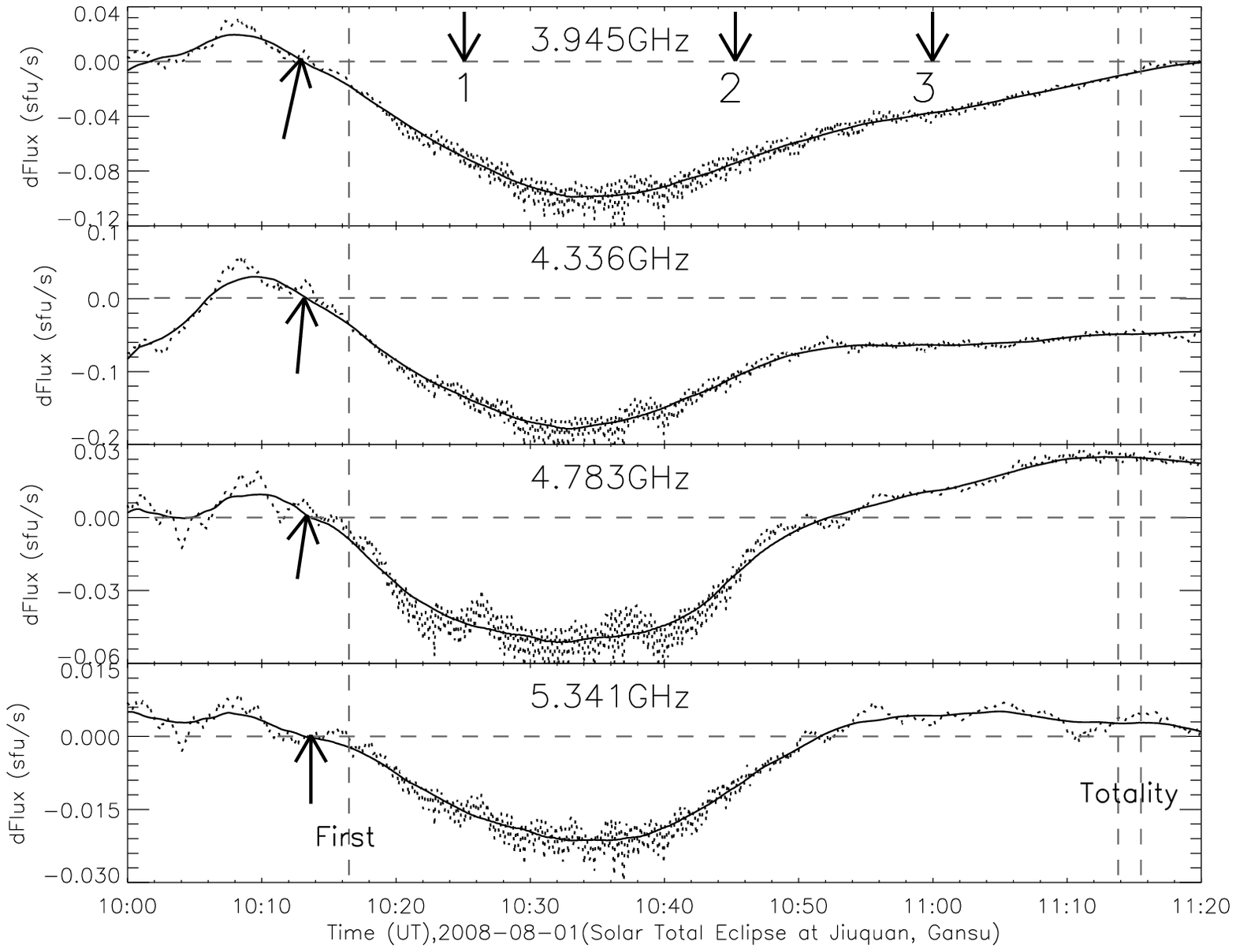}
  \includegraphics[width=5.4cm]{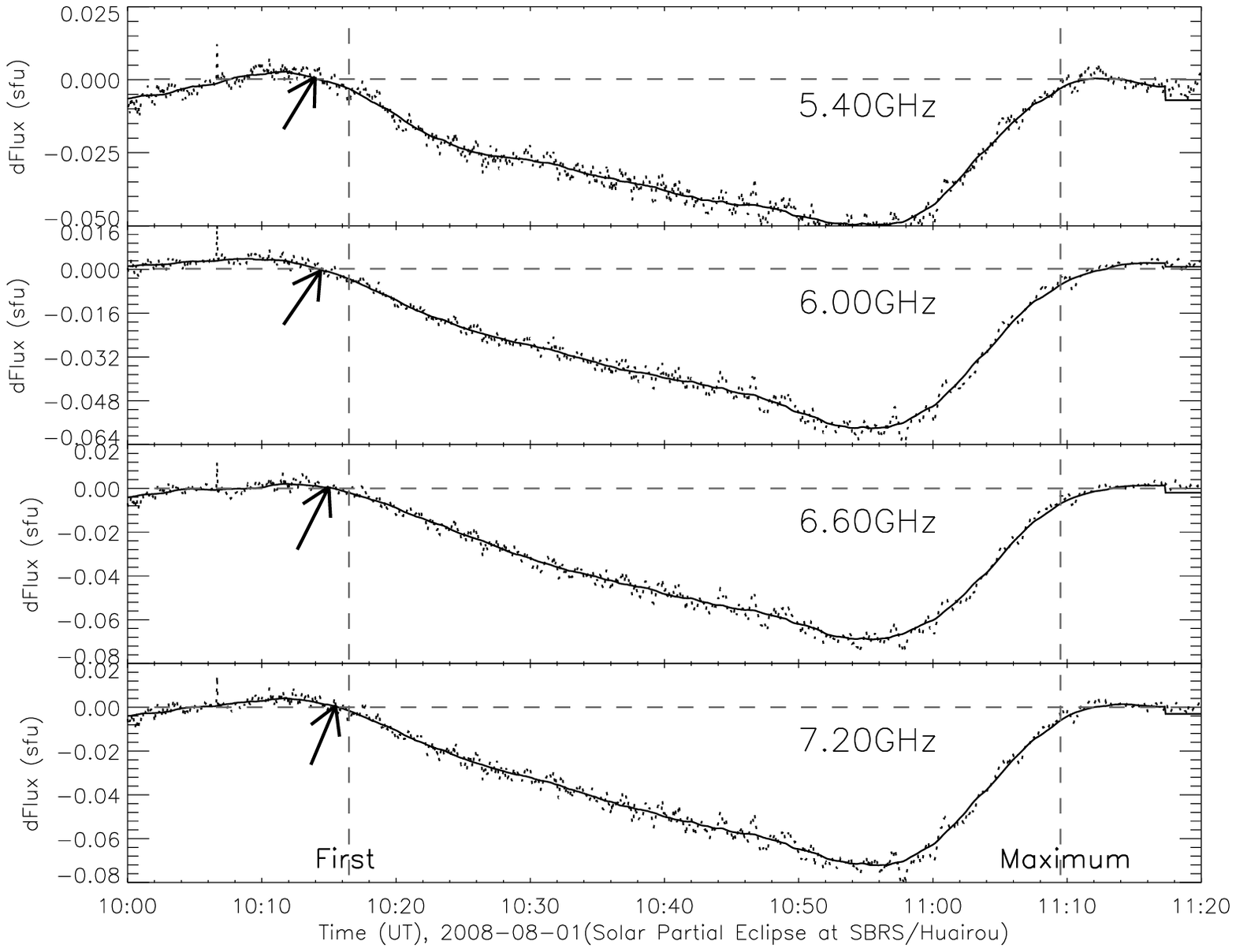}
   \caption{Solar radio eclipse slope curves of August 1, 2008. Left: 2.60 -- 3.80 GHz, partial eclipse, SBRS/Huairou,
   Beijing; Middle: 3.80 -- 5.20 GHz, total eclipse, Jiuquan, Gansu; Right: 5.20 -- 7.60 GHz, partial eclipse, SBRS/Huairou, Beijing.
   The arrows below the curves indicate locations of the first solar radio contact of the eclipse, and the vertical arrows with number
   1, 2, 3 indicate three instants as in Figure 1.}
\end{center}
\end{figure}

\section{Results and some Implications about Quiet Sun
Atmospheric Model}

\subsection{Results}

Generally, from the solar radio eclipse curves we can get the time
of the first radio contact and determinate the solar radio
radii$^{[7]}$. However, from Figure 1 we can find that the
fluctuation of the radio eclipse curves is fairly strong. It is
not easy to determine the exact time of the first radio contact of
the eclipse just from the solar radio eclipse curves (dashed lines
in Figure 1), even if from the smoothed curves (solid thick lines
in Figure 1). Here, we apply the solar radio eclipse slope curves
(Figure 2) to obtain the relatively exact time of the first solar
radio contact. The following is the basic working principle: the
telescope antenna is always pointed to the center of the solar
disk, before the first radio contact. The receiver gets the whole
solar radio emission, and the slope curves should be approximated
to a horizontal line around zero; from the beginning of the
eclipse (first contact), when the receiver gets less and less
solar radio emission with the increasing of the solar disk
occulted by the Moon, the slope curve will become negative. Then
we may define the intersection point between the smoothed slope
curve and the zero horizontal line as the time of the first solar
radio contact of the eclipse. The arrows in Figure 2 indicate the
time of the first solar radio contact of the eclipse at each
frequency, with corresponding values ($t_{ra}$) listed in Table 1.

Based on $t_{ra}$, we can calculate the solar radio radii ($r$):

\begin{equation}
r=\frac{t_{1}-t_{ra}}{t_{\odot}}+1.
\end{equation}

Here, $t_{1}$ is the time of the first optical contact of the
eclipse, as for the solar total eclipse at Jiuquan,
$t_{1}$=10$^{h}$16$^{m}$36.5$^{s}$ UT; as for the partial eclipse
at Huairou, Beijing, $t_{1}$=10$^{h}$16$^{m}$36.1$^{s}$ UT.
$t_{\odot}$ indicates the time when the Moon's limb moves for a
distance of a solar photospheric radius $R_{\odot}$, and can be
calculated from the following equation:

\begin{equation}
t_{\odot}=\frac{t_{max}-t_{1}}{\frac{\sqrt{(R_{\odot}+r_{mo})^{2}-h^{2}}}{R_{\odot}}}
\end{equation}

$t_{max}$ stands for the time of the maximum phase of the eclipse,
$r_{mo}$ for the lunar radius at the time of the first optical
contact, and $h$ for the distance between the solar disk center
and the lunar center at the maximum phase. As for the solar total
eclipse at Jiuquan, $t_{max}$=11$^{h}$14$^{m}$31.7$^{s}$ UT,
$h=12.95$ arcsecond, $r_{mo}=978.1$ arcsecond, then
$t_{\odot}=1705.18$ s; as for the partial eclipse at SBRS/Huairou
$t_{max}$=11$^{h}$09$^{m}$23.2$^{s}$ UT, $h=190.04$ arcsecond,
$r_{mo}=977.7$ arcsecond, then $t_{\odot}=1564.70$ s. Here, we
assume that the relative velocity of the Moon versus the Sun is
uniform.

\begin{figure}
\begin{center}
  \includegraphics[width=10.0cm]{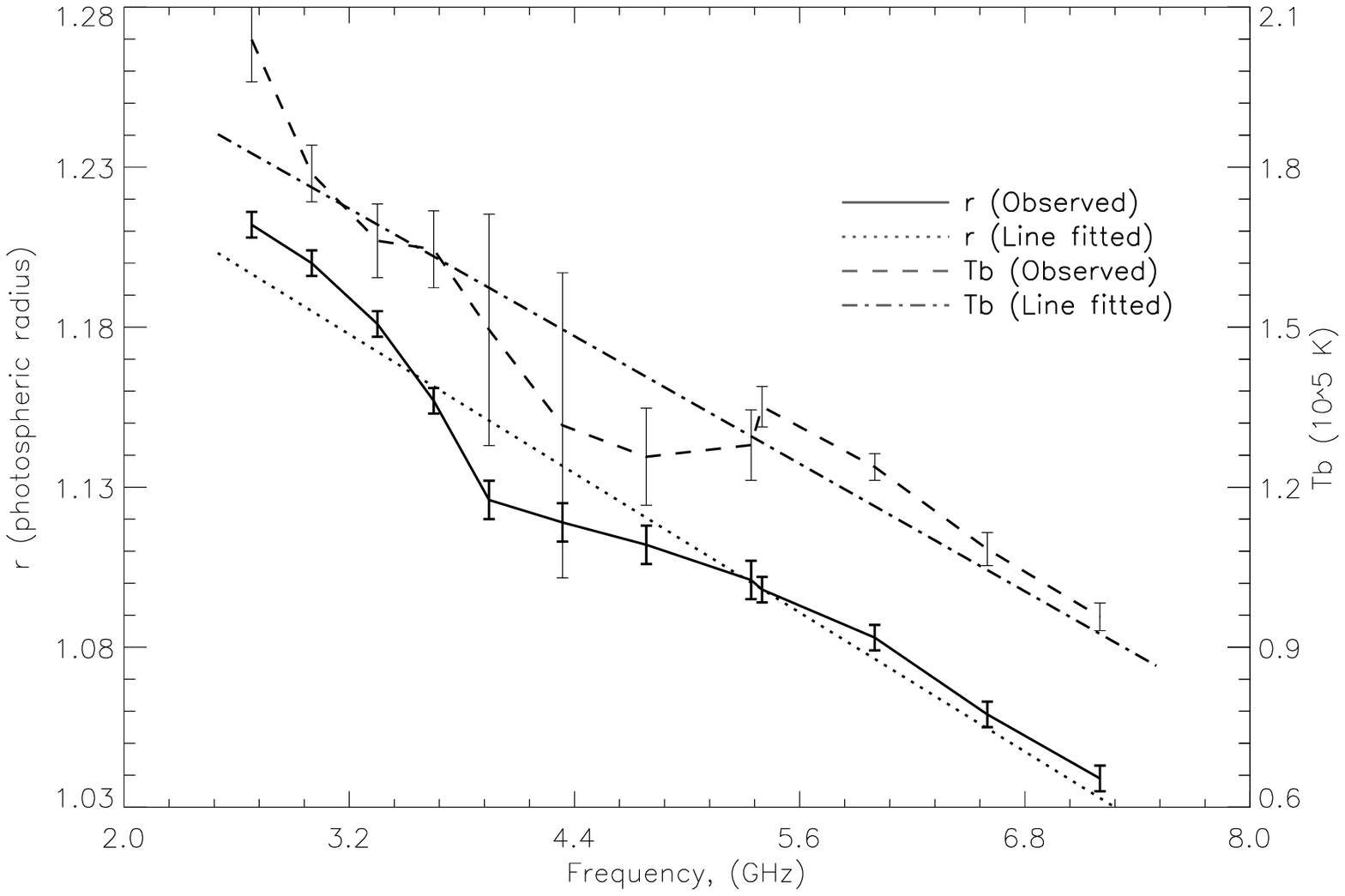}
  \caption{Results of the solar radio radii ($r$, in unit of solar photospheric radius) and the bright temperature ($T_{b}$) in
  the frequency of 2.0 -- 7.60 GHz observed in the solar eclipse of 2008-08-01 at EcBS/Jiuquan and SBRS/Huairou.}
\end{center}
\end{figure}

\begin{table}\def~{\hphantom{0}}
  \begin{center}
  \caption{The list of VLP, LPP and SPP occurred in the flare event on 2006-12-13. Here the frequency drifting rate is the averages values in the
  unit of MHz/s, and the bandwidth is also in MHz, both of duration and period are in unit of second.}
  \label{tab:kd}
  \begin{tabular}{lccccccccccccc}\hline
  $ f   $(GHz)&      $t_{ra}$ (UT)       & $r$(R$_{\odot}$)& $   Flx $ (sfu) & $T_{b}(10^{5}K)$  &$n_{e}(10^{9}cm^{-3})$& Instrument    \\\hline
 $2.680$     & 10$^{h}$11$^{m}$04$^{s}$ & $1.212\pm0.002$ & $71.5\pm2.72  $ & $2.039\pm0.079$    &  $2.06\pm0.05$     & SBRS/Huairou \\
 $3.000$     & 10$^{h}$11$^{m}$22$^{s}$ & $1.200\pm0.002$ & $77.0\pm2.32  $ & $1.788\pm0.053$    &  $2.11\pm0.04$     & SBRS/Huairou \\
 $3.350$     & 10$^{h}$11$^{m}$53$^{s}$ & $1.181\pm0.002$ & $86.5\pm3.55  $ & $1.662\pm0.069$    &  $2.25\pm0.05$     & SBRS/Huairou \\
 $3.650$     & 10$^{h}$12$^{m}$30$^{s}$ & $1.157\pm0.002$ & $97.5\pm4.24  $ & $1.646\pm0.072$    &  $2.43\pm0.06$     & SBRS/Huairou \\
 $3.945$     & 10$^{h}$12$^{m}$01$^{s}$ & $1.126\pm0.003$ & $98.0\pm13.94 $ & $1.495\pm0.217$    &  $2.46\pm0.20$     & EcBS/Jiuquan \\
 $4.336$     & 10$^{h}$12$^{m}$13$^{s}$ & $1.119\pm0.003$ & $103.0\pm22.38$ & $1.316\pm0.286$    &  $2.49\pm0.31$     & EcBS/Jiuquan \\
 $4.783$     & 10$^{h}$13$^{m}$25$^{s}$ & $1.112\pm0.003$ & $118.0\pm8.50 $ & $1.257\pm0.091$    &  $2.67\pm0.12$     & EcBS/Jiuquan \\
 $5.341$     & 10$^{h}$13$^{m}$44$^{s}$ & $1.101\pm0.003$ & $147.0\pm7.45 $ & $1.279\pm0.066$    &  $3.02\pm0.09$     & EcBS/Jiuquan \\
 $5.400$     & 10$^{h}$14$^{m}$02$^{s}$ & $1.098\pm0.002$ & $158.0\pm4.36 $ & $1.351\pm0.038$    &  $3.18\pm0.05$     & SBRS/Huairou \\
 $6.000$     & 10$^{h}$14$^{m}$27$^{s}$ & $1.083\pm0.002$ & $173.6\pm3.33 $ & $1.238\pm0.025$    &  $3.33\pm0.04$     & SBRS/Huairou \\
 $6.600$     & 10$^{h}$15$^{m}$03$^{s}$ & $1.059\pm0.002$ & $176.2\pm5.32 $ & $1.084\pm0.031$    &  $3.35\pm0.06$     & SBRS/Huairou \\
 $7.200$     & 10$^{h}$15$^{m}$34$^{s}$ & $1.039\pm0.002$ & $178.0\pm4.95 $ & $0.957\pm0.026$    &  $3.37\pm0.06$     & SBRS/Huairou \\\hline\hline
  \end{tabular}
 \end{center}
\end{table}

The observational solar radio radii and the emission flux before
the time of the first radio contact of the eclipse are also listed
in Table 1. Figure 3 presents the distribution of the solar radio
radii with frequencies. From this figure we find that the curve is
most likely to be a beeline. We operate a best linear fit and get
a linear function:

\begin{equation}
r=1.29370-0.036224f
\end{equation}

Here, the unit of solar radio radius $r$ is in solar photospheric
radius ($R_{\odot}$), and the frequency $f$ is in GHz.

Then, we calculate the bright temperature of the radio quiet Sun
at each frequency:

\begin{equation}
T_{b}=\frac{c^{2}D^{2}Flx}{k_{B}f^{2}(rR_{\odot})^{2}}=3.009\times10^{22}\frac{Flx}{f^{2}r^{2}}
\end{equation}

Here, $Flx$ is unit in sfu, and $f$ is in Hz. $D$ is the averaged
distance from the Sun to Earth. The bright temperature at each
frequency is also listed in Table 1, and plotted in Figure 3.

As mentioned in section 1, the Sun was in its quiet phase. We may
simply assume that: (1) the microwave emission is mainly
originated from the bremsstrahlung mechanism of thermal plasma in
which the electron density is approximated to that of ions
($n_{e}\approx n_{i}$); (2) the solar atmosphere is spherically
symmetrical; (3) the absorption effects can be ignored. Then we
can make out the estimation of the electron plasma density:

\begin{equation}
n_{e}\simeq
1.2416\times10^{12}\frac{Flx^{1/2}T_{b}^{1/4}}{r(M\triangle l)
^{1/2}},(cm^{-3})
\end{equation}

Here, $M=19.38+\frac{3}{2}ln T_{b}-ln f$, and $\triangle l$ is the
characteristic length of the emitting medium along the line of
sight. Simply, we make $\triangle l\sim 5000$ km for all
frequencies. The result of $n_{e}$ is listed in Table 1, and
plotted in Figure 4.

From these results we find that the bright temperature of the
coronal plasma is in the range of $9.57\times10^{4}$ --
$2.04\times10^{5}$ K, and the plasma density is in the range of
$2.05\times10^{9}$ -- $3.37\times10^{9}cm^{-3}$ at the height of
$2.7\times10^{4}$ -- $1.5\times10^{5}$ km ($r=1.039 -
1.212R_{\odot}$) above the solar photosphere.

\subsection{Error Analysis}

Generally, the fluctuation of the radio emission during the
eclipse was fairly strong with obvious errors, so was the case
with Huairou and Jiuquan observations, because SBRS was old and
ECBS was new. We make a broad averaging with the data. The error
of solar radio radii ($r$), bright temperature ($T_{b}$), and
electron plasma density ($n_{e}$) can be expressed as: $\triangle
r=\pm\frac{\triangle t_{ra}}{t_{\odot}}$, $\triangle T_{b}=\pm
T_{b}(\frac{\triangle Flx}{Flx}+\frac{2\triangle
f}{f}+\frac{2\triangle r}{r})$, and $\triangle n_{e}=\pm
n_{e}(\frac{\triangle r}{r}+\frac{\triangle
Flx}{2Flx}+\frac{\sqrt{M}-3}{4M}\frac{\triangle T_{b}}{T_{b}})$,
respectively. We make 3 times of standard deviation as the error
of emission flux, $\triangle Flx= 3\sigma_{Flx}$. As for the data
of EcBS/Jiuquan, we make averaging in each recorded file, so the
integrated time is about 3 -- 7 s with the mean value of 5.59 s,
so $\triangle t_{ra}=5.59$s, $\triangle r=\pm0.003 R_{\odot}$;
$\triangle f=28$ MHz. As for the data of SBRS/Huairou, we make
averaging in each segment of 3.2 s, $\triangle t_{ra}=3.2$ s,
$\triangle r=\pm0.002 R_{\odot}$; $\triangle f=10$ MHz in
frequency of 2.60--3.80 GHz, and 20 MHz in 5.20--7.60 GHz. Then
$\triangle T_{b}$, and $\triangle n_{e}$ can be made out (listed
in Table 1).

These analyses show that the error level of EcBS/Jiuquan is higher
than that of SBRS/Huairou.

\subsection{Implications about Quiet Sun Atmospheric Model}

The solar radio radius is an important quantity in studying the
solar atmospheric model since the height above the photosphere at
which the radio emission originates at different frequencies
defines the plasma density distribution. With these results, we
can deduce the seimiempirical solar atmospheric model. The optical
and millimeter radio observations can present the solar
photospheric and chromospheric models$^{[8,9,10,11]}$. Fontenla,
Balasubramniam, and Harder gave a semiempirical model of the quiet
Sun low chromosphere at moderate resolution$^{[12]}$. The cm and
dm wavelength radio observations will provide the information of
the solar coronal atmospheric model. From the above results, we
know that the emission region of the 2.60-7.60 GHz radio emission
must be located in the corona. As most of the eruptive processes
always occur in these region, a perfect coronal model, especially
the semiempirical models from the observations, is very important
for understanding physical processes in the Sun. During the quiet
phase of solar cycle, the magnetic field was very weak and fairly
uniform. There was almost no magnetic active regions. It is easier
to set up a fairly perfect atmospheric model of the quiet Sun.

\begin{figure}
\begin{center}
  \includegraphics[width=10.0cm]{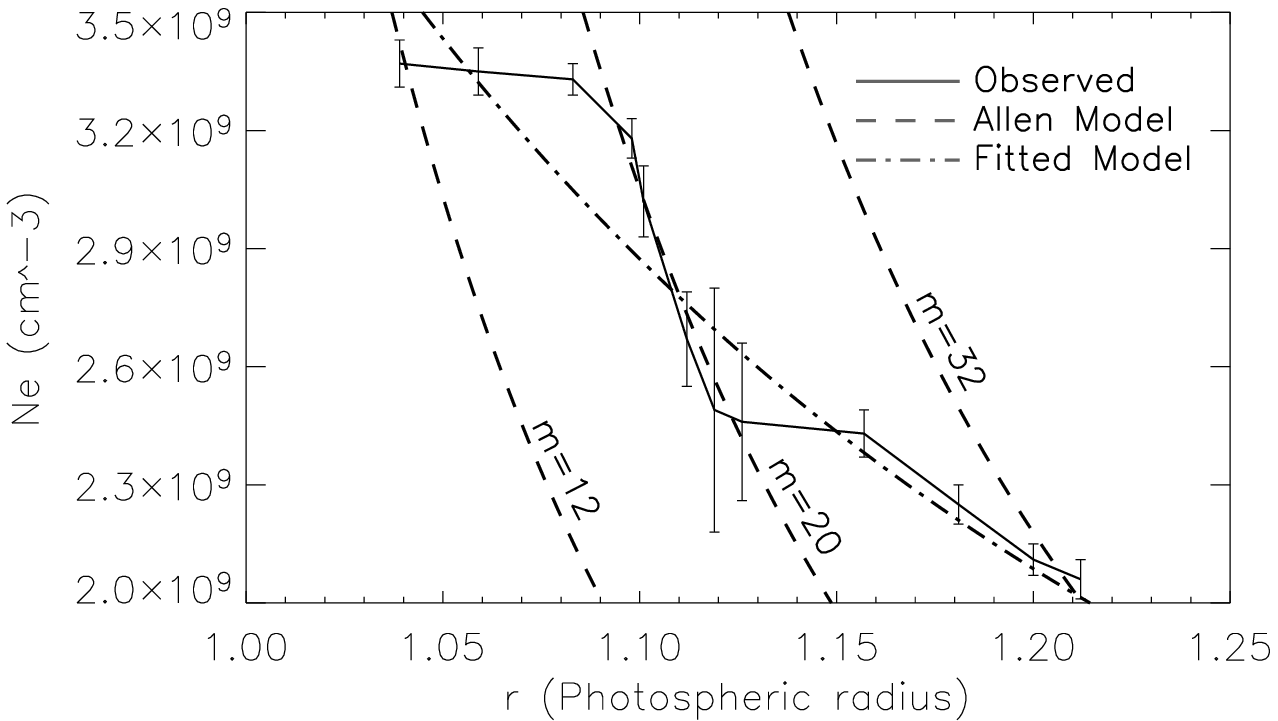}
  \caption{A Comparison between the observed result of coronal density and the Allen's model. The solid line represents
  the observed result. The dashed lines are Allen's models with $m$ times. The dashed-dotted line is a best-fitted model expressed in Equation (6).}
\end{center}
\end{figure}

Figure 4 presents the comparison between the observed result of
coronal density and the classic Baumbach-Allen's model.
Baumbach-Allen's model$^{[13]}$ can be expressed as:

\begin{equation}
n_{e}=m\times10^{8}(\frac{1.55}{r^{6}}+\frac{2.99}{r^{16}}),(cm^{-3})
\end{equation}

Here $m$ is a given multiple. The solid line stands for the
observed result. The dashed lines are Allen's models with $m$
times. Here $m$ is in the range of 12 -- 32. However, from this
figure we find that the curves of Baumbach-Allen's model are too
steep to fit the observed results. We make a best-fit curve, the
dashed-dotted line in Figure 4 is a best-fitted model which can be
expressed as follows:

\begin{equation}
n_{e}\simeq
1.42\times10^{9}(\frac{1}{r^{2}}+\frac{1.93}{r^{5}}),(cm^{-3})
\end{equation}

Based on the comparison between equations (6) and (7), we find
that coronal plasma density is more likely to be associated with
the low power of height ($r$) than that of the classic model.

\section{Discussions and Conclusions}

In this study work, we analyze the joint-observations of radio
broadband spectral emissions during the solar eclipse on August 1,
2008 at Jiuquan (total eclipse, observed in the frequency of 2.00
-- 5.60 GHz) and Huairou (partial eclipse, observed in the
frequency of 2.60 -- 3.80 GHz and 5.20 -- 7.60 GHz). Using these
telescopes, we can assemble a successive series of broadband
spectrum with frequency of 2.60 -- 7.60 GHz to observe the solar
eclipse synchronously. With these analyses, we obtain a
semiempirical quiet Sun coronal model (equation (7)) which
indicates that the distribution of solar coronal plasma density in
respect to the height is not so steep as the classic model,
corresponding to the height of $2.7\times10^{4}$ --
$1.5\times10^{5}$ km ($r=1.039 - 1.212R_{\odot}$) above the solar
photosphere, and the bright temperature is in the range of
$9.57\times10^{4} - 2.04\times10^{5}$ K. This is consistent with
the result of Ref [14].

However, our results are obtained from the range of 2.60 -- 7.60
GHz, and only valid in the limited space of $r=1.039 -
1.212R_{\odot}$. It is necessary to enlarge the observing
frequency range in the future observations. Additionally, the
height angle of the eclipse of August 1, 2008 was very small. When
the eclipse occurred, the line-of-sight was fairly close to the
ground, with many disturbances coming from the subaerial factors.
The telescope of EcBS/Jiuquan, a new instrument, needs to be
improved in many aspects. Fortunately, a more significant solar
total eclipse will occur on July 22, 2009 in a more extended area
along the Changjiang River, during 8-11 o'clock in the morning,
and the height angle is very high. The observational condition is
very good. We will make more detailed observations to get the
further supports for our semiempirical model.

\acknowledgments

The authors would like to thank the anonymous referee for helpful
and important comments. They also thank associate professor Cheng
Zhuo from Purple Mountain Observatory of CAS (Nanjing) for the
important data of the optical solar eclipse tracks of August 1,
2008. This work was supported by the CAS-NSFC Key Project (Grant
No.10778605), the National Science Foundation of China (Grant
Nos.10733020, 10843002, 10873021), and the MOST (Grant
No.2006CB806301).

\end{document}